\documentclass{jetpl}
\usepackage{amsfonts}
\usepackage{amssymb}
\usepackage{amsthm} 
\usepackage{newlfont}
\usepackage{epsfig}
\usepackage{amscd}
\usepackage{graphicx}
\usepackage{epsfig}
\usepackage{color}
\twocolumn

\lat

\title{Asymmetry of bipartite quantum discord}

\rtitle{Asymmetry of bipartite quantum discord}

\sodtitle{Asymmetry of bipartite quantum discord}

 \author{
E.\,B.\,Fel'dman$^+$, A.\,I.\,Zenchuk $^{+}$
}

\rauthor{
E.\,B.\,Fel'dman, A.\,I.\,Zenchuk \/\thanks{e-mail: zenchuk@itp.ac.ru}
}

\sodauthor{Fel'dman, Zenchuk }

\address{$^+$Institute of Problems of Chemical Physics, Russian Academy of Sciences,
Chernogolovka, Moscow reg., 142432, Russia}

\dates{}

\abstract{
It is known from the analysis of the density matrix for bipartite systems that the quantum discord (as a measure of quantum correlations) depends on the particular subsystem chosen for the projective measurements. 
We study    asymmetry of the discord in a simple  physical model of two spin-1/2 particles with the dipole-dipole interaction governed by the XY Hamiltonian in the inhomogeneous magnetic field.
The dependence of the above discord asymmetry   on the Larmour frequencies at both $T=0$ (the ground state) and  $T>0$  has been investigated.
It is demonstrated, in particular, that the asymmetry is negligible for high temperatures but it may become significant with the  decrease in temperature. 
}


\begin{document}

\maketitle

\section{Introduction}
The effective manipulations by quantum correlations results in significant  advantages of the quantum devices (in particular, quantum computers) in comparison  with their classical counterparts.  However, the problem of correct measure of  quantum correlations has not been  resolved up to now.
Till recently,  the entanglement has been used  
  as a measure of  quantum correlations \cite{W,HW,P,AFOV,DPF}. 
 However, 
it was shown both theoretically \cite{BCJLPS,M}
 and experimentally 
\cite{LBAW} that some mixed separable states (i.e. states with zero entanglement) 
 allow one to realize the advantages of  quantum algorithms  in comparison with their classical analogies and quantum nonlocality
has been observed in the systems  without entanglement  \cite{BDFMRSSW}. 
 Such observations suggest us to conclude that the entanglement does not involve  all quantum correlations responsible for the advantages of  quantum algorithms in comparison with the classical ones. 

The concept of 
 quantum discord, which is intensively developing during last years \cite{OZ,L,ARA}, is based on a separation of a quantum part out of the total  mutual information encoded into a bipartite system. 
The discord is completely defined by the quantum properties of the system and  becomes zero for the classical systems.  

The evaluation of the quantum discord is a cumbersome computational problem so that the methods of its
analytical  calculations have been developed only for two-qubit systems \cite{L,ARA}.  Calculation of the discord 
consists of two steps: (a) calculation of the  mutual  information encoded into two subsystems $A$ and $B$ of the bipartite  quantum system  and (b) calculation of the classical component of this mutual information.
The second part is the most cumbersome one because it is based on the multiparameter optimization over the von Neumann type of measurements performed on one of the subsystems (say, $B$). 
It is seemed out that the quantum discord depends on which subsystem is taken for  
measurements { \cite{OZ}.}
{  It will be shown below that the difference between the quantum discords calculated using the projective measurements performed over the  different subsystems of a bipartite system  (the asymmetry) can be large (about 25$\%$). This means that the quantum discord as a measure of quantum correlations does not provide an unambiguous quantitative characteristic of these correlations. As a result, the quantum discord is insufficient for an estimation of the utility  of different materials in quantum devices and requires modification.  A possible modification of the quantum discord is suggested in this letter.  } 

The conditions needed to provide the symmetry of the discord with respect to subsystems $A$ and $B$ 
have been found only for the particular case of the two-qubit system 
\cite{ARA}. These conditions are  based on the analysis of the structure of the two-qubit density matrix and have not been related with the physical parameters of the system (such as a temperature and Larmour frequencies). 
{ A connection of the quantum discord with the real physical parameters opens a direct way for the experimental investigations of the discord and its asymmetry  and allows one to control the quantum discord by means of physical parameters.}

This letter is devoted to the problem of asymmetry of the  quantum descord of the two-qubit  system with respect to the subsystems $A$ and $B$ chosen for the projective measurements. Considering the simple physical model of two spin-1/2 particles with the dipole-dipole interaction governed by the XY Hamiltonian in the inhomogeneous magnetic field we study the dependence of the above asymmetry on the temperature and Larmour frequencies.

\section{Model} 
\label{Section:model}
 We consider the quantum system of two spin-1/2 particles governed by   the XY  Hamiltonian  in the  external inhomogeneous magnetic field.  The Hamiltonian of the system reads:
\begin{multline}
\label{XY}
H={D} (I_{1,x} I_{2,x}+I_{1,y} I_{2,y} ) +D\Omega (1+ \Delta) I_{1,z} +\\
 D\Omega(1- \Delta) I_{2,z},
\end{multline}
where $D$ is the constant of the dipole-dipole interaction, $I_{n,\alpha}$ ($n=1,2$, $\alpha=x,y,z$)- $n$th spin projection operator on the axis $\alpha$, $D\Omega(1\pm\Delta)$ are the Larmour frequencies.
We assume that $D>0$, $0\le \Delta\le 1$ and $\Omega\ge 0$.
The system is 
in the  thermal equilibrium state with the Gibbs density matrix:
\begin{eqnarray}\label{rho}
\rho = e^{-\beta H} /Z,\;\;Z={\mbox{Tr}}\; e^{-\beta H},\;\;\beta = \frac{\hbar}{kT},
\end{eqnarray}
where  $T$ is the temperature, $k$ is the Boltsman constant. 
It is simple to check that this  density matrix has the following form:
\begin{eqnarray}\label{rhog}
\rho=\left(\begin{array}{cccc}
\rho_{11}&0&0&0\cr
0&\rho_{22}&\rho_{23}&0\cr
 0&\rho_{23}&\rho_{33}&0\cr
 0&0&0&\rho_{44}
\end{array}
\right),
\end{eqnarray}
where all entries are real numbers.
The density matrix (\ref{rhog}) can be considered as 
a particular case of  so-called two-qubit X-matrix \cite{ARA}. The quantum discord for such matrices can be calculated by the method developed in   \cite{ARA}, { where optimization over three continues parameters is reduced to the calculation of the minimum  of six values}. This method 
becomes much simpler in our  case. In particular, the optimization must be performed over the single parameter (instead of three parameters in ref.\cite{ARA}). { As a result, the formulas for the discord calculation becomes simpler and the optimization is reduced to the calculation of the minimum  of two values}.

Assuming that the von Neumann type measurements are performed over the subsystem $B$, one can define quantum discord $Q^B$ as follows \cite{OZ}: 
\begin{eqnarray}\label{Q}
Q^B={\cal{I}}(\rho) -{\cal{C}}^B (\rho).
\end{eqnarray}
Here   ${\cal{I}}(\rho)$ is the total mutual information \cite{OZ} which may be written as follows: 
\begin{eqnarray}\label{I}
&&
{\cal{I}}(\rho) =S(\rho^A) + S(\rho^B) + \sum_{j=0}^3 \lambda_j \log_2 \lambda_j,\\\nonumber
\end{eqnarray}
where  $\lambda_j$ ($j=0,1,2,3$) are eigenvalues of the density matrix $\rho$, $\rho^A={\mbox{Tr}}_B \rho$ and $\rho^B={\mbox{Tr}}_A \rho$ are the reduced density matrices
and the appropriate entropies $S(\rho^A)$ and $S(\rho^B)$ are given by the following formulas:
\begin{align}\label{SAB}\nonumber
S(\rho^A){}&=-(\rho_{11} + \rho_{22} ) \log_2(\rho_{11}+\rho_{22}) -\\\nonumber
           &(\rho_{33} + \rho_{44} ) \log_2(\rho_{33}+\rho_{44}) ,\\\nonumber
S(\rho^B){}&=-(\rho_{11} + \rho_{33} ) \log_2(\rho_{11}+\rho_{33}) -\\
         &(\rho_{22} + \rho_{44} ) \log_2(\rho_{22}+\rho_{44}) .
\end{align}
The so-called classical counterpart ${\cal{C}}^B (\rho)$ of the mutual information 
can be found considering the minimization over projective measurements on the subsystem $B$ as follows \cite{ARA}:
\begin{eqnarray}\label{CB2}
&&
{\cal{C}}^B (\rho)=S(\rho^A) -\min\limits_{\eta=(0,1)}(p_0 S_0 + p_1 S_1),
\end{eqnarray}
where
\begin{align}\label{S}
&S_i = -\frac{1-\theta^{(i)}}{2}\log_2\frac{1-\theta^{(i)}}{2}-
\\\nonumber
&\hspace{1cm}
                 \frac{1+\theta^{(i)}}{2}\log_2\frac{1+\theta^{(i)}}{2},
\\\label{p}
&p_i=\frac{1}{2} \Big(1+(-1)^i\eta(2(\rho_{11}+\rho_{33})-1) \Big),\\\label{theta}
&\theta^{(i)}=\frac{1}{p_i}\Big[(1-\eta^2) \rho_{23}^2+\\\nonumber
&
\frac{1}{2}
\Big(
2(\rho_{11}+\rho_{22})-1 +(-1)^i \eta(1-2(\rho_{22}+\rho_{33}))\Big)^2 
 \Big]^{1/2},
\\\nonumber
&
i=0,1.
\end{align}
It is simple to show that the quantum discord $Q^A$ obtained performing 
 the von Neumann type measurements  on the subsystem $A$ can be calculated 
as follows:
\begin{eqnarray}\label{QA}
Q^A=Q^B|_{\rho^{(22)}\leftrightarrow \rho^{(33)}}
\end{eqnarray}
for the system with the density matrix $\rho$ given by eq.(\ref{rhog}).

\section{The quantum discord of the ground state}
\label{Section:gr}
In the considered model, the ground state is defined by the minimal eigenvalue of the Hamiltonian. 
Since the eigenvalues read
\begin{align}
\lambda^{(1)}&=-D\Omega,\;\;\;
\lambda^{(2)}=D\Omega ,\\\nonumber
\lambda^{(3)}&=-\frac{D}{2}\sqrt{1+4\Omega^2\Delta^2},\\\nonumber
\lambda^{(4)}&=\frac{D}{2}\sqrt{1+4\Omega^2\Delta^2},
\end{align}
one can conclude  that the minimal eigenvalue $\lambda_{min}$ is either $\lambda^{(1)}$ or $\lambda^{(3)}$ depending on the values  $\Delta$ and $\Omega$. For the fixed $\Delta$, we introduce the    critical value $\Omega_c$ such that $\lambda^{(1)}=
\lambda^{(3)}$ at $\Omega=\Omega_c$, so that $\lambda^{min}=\lambda^{(3)}$ if $\Omega<\Omega_c$ and $\lambda^{min}=\lambda^{(1)}$ if $\Omega>\Omega_c$. 

\begin{figure*}
   \epsfig{file=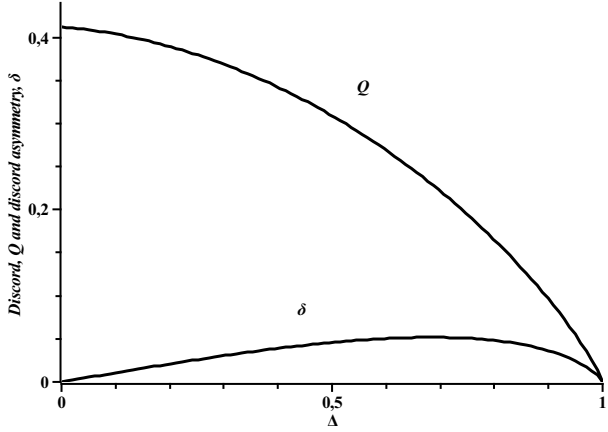
   , scale=0.5,angle=270
}
  \caption{Fig.1. The discord asymmetry  $\delta=Q^A-Q^B$ and the discord  $Q$ as  functions of $\Delta$ at $\Omega=\Omega_c$.  }
  \label{Fig:gr} 
\end{figure*}

The critical value $\Omega_c$  is following:
\begin{eqnarray}\label{Omegac}
\Omega_c=
\frac{1}{2\sqrt{1- \Delta^2}}
\end{eqnarray}
Both the calculation of the discord $Q^{B}$ using formulas (\ref{Q}-\ref{theta})  and the calculation of the discord  $Q^{A}$ using eq.(\ref{QA}) demonstrate that $Q^A=Q^B$ if $\Omega<\Omega_c$ and $\Omega>\Omega_c$, i.e. there is no asymmetry. The discord  asymmetry $\delta=Q^A-Q^B$ appears at $\Omega=\Omega_c$ and equals to 
\begin{align}\nonumber
\delta(\Delta)&=\frac{1}{4} \left(
3 \log_2\frac{3-\Delta}{3+\Delta}-\log_2\frac{1+\Delta}{1-\Delta}-\Delta
\log_2\frac{9-\Delta^2}{1-\Delta^2} +\right.
\\\nonumber
&
\left.\sqrt{2(1+\Delta)}\log_2\frac{2+\sqrt{2(1+\Delta)}}{2-\sqrt{2(1+\Delta)}} +\right.
\\\label{delta}
&\left.
\sqrt{2(1-\Delta)}\log_2\frac{2-\sqrt{2(1-\Delta)}}{2+\sqrt{2(1-\Delta)}} 
\right),
\end{align}
whereas
\begin{align}\label{AB}
Q^{B,A}_c(\Delta)&=
\frac{1}{4} \Big( 12 -(3\mp\Delta)\log_2(3\mp\Delta) -\\\nonumber
&(1\pm\Delta)\log_2(1\pm\Delta)-\\\nonumber
&
(2-\sqrt{2(1\pm\Delta)}) \log_2(2-\sqrt{2(1\pm\Delta)}) -\\\nonumber
&
(2+\sqrt{2(1\pm\Delta)}) \log_2(2+\sqrt{2(1\pm\Delta)})\Big).
\end{align}
Thus, if $\delta\neq 0$, then it seems to be reasonable to define the discord of a bipartite system by the following formula:
\begin{eqnarray}
Q=\min(Q^A,Q^B).
\end{eqnarray}
It follows from eq.(\ref{delta}) that the  maximal asymmetry $\delta\approx 0.052$ 
is achieved  at $\Delta_{max}\approx 0.683$ when $\Omega_c\approx 0.684$. Appropriate values of the discords are  $Q^{B}_c\approx 0.230$, $Q^A_c\approx 0.282$, { i.e. the asymmetry is about 23$\%$}. 
The dependence of both the discord asymmetry $\delta$ and the discord $Q$ on the parameter  $\Delta$
is represented in Fig.\ref{Fig:gr}. 

{  The absence of the asymmetry for both $\Omega<\Omega_c$ and $\Omega>\Omega_c$ is readily explained by the fact that the ground state is pure unless $\Omega=\Omega_c$. It is known that  the discord of a pure state equals to the entanglement and does not depend on which subsystem is taken for the projective measurements \cite{DSC}. On the contrary, the ground state is degenerated if $\Omega=\Omega_c$, so that the state becomes mixed which (together with the asymmetry of the quantum system)  results in  the discord asymmetry. If $\Delta= 0$, then the system becomes symmetrical so that $\delta=0$, see Fig.\ref{Fig:gr}.  If $\Delta=1$, then $\lambda_1>\lambda_3$ so that the ground state is  pure for all $\Omega$ and, consequently,  $\delta=0$.  One has to emphasize also that the ground state can be obtained as the limit of density matrix (\ref{rho}) when the temperature $T$ tends to zero, which uniquely yields either the pure (nondegenerate) or mixed (degenerate) ground state depending on whether $\Omega\neq \Omega_c$ or $\Omega=\Omega_c$. 
 }







\section{The thermal quantum discord}
\begin{figure*}
   \epsfig{file=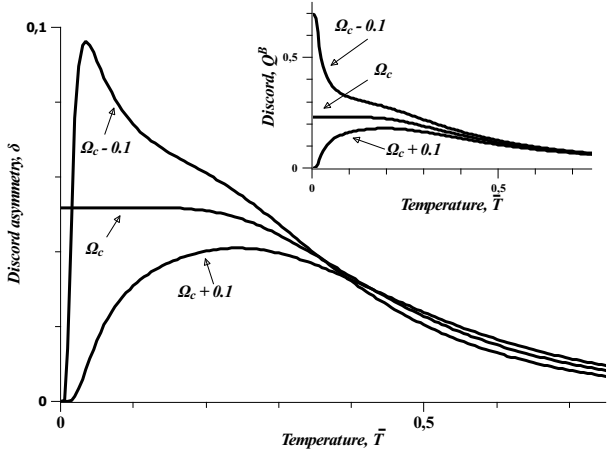
   , scale=0.5,angle=270
}
  \caption{Fig.2. The  discord asymmetry  $\delta=Q^A-Q^B$ and the  discord  $Q$  (inset) as  functions of the dimensionless  temperature $\bar T=kT/(\hbar D)$ for $\Delta=\Delta_{max}\approx 0.683$ and different $\Omega$: 
$\Omega=\Omega_c(\Delta_{max})-0.1;\;\Omega_c(\Delta_{max});\;\Omega_c(\Delta_{max})+0.1$. 
 }
  \label{Fig:Heis} 
\end{figure*}

The model represented in Sec.\ref{Section:model}~~
allows one to investigate the asymmetry of the thermal discord at $T>0$. We study it for the case  $\Delta=\Delta_{max}\approx 0.683$  with the different values of the parameter  $\Omega$: $\Omega=\Omega_c(\Delta_{max})-0.1;\;\Omega_c(\Delta_{max});\;\Omega_c(\Delta_{max})+0.1$.
Results of such calculations  are represented in Fig.\ref{Fig:Heis}. 
The asymmetry of the thermal discord is negligible at high temperatures when quantum correlations disappear. However, the asymmetry increases with the decrease in 
the temperature and becomes considerable over a  wide temperature range. Asymmetry disappears with $T\to 0$ for all $\Omega$  except  $\Omega=\Omega_c$ when $\delta\to  0.052$ and $Q\to 0.230$, { i.e. the asymmetry is about $ 23 \%$ which agrees with the results of Sec.\ref{Section:gr}.}

\begin{figure*}
   \epsfig{file=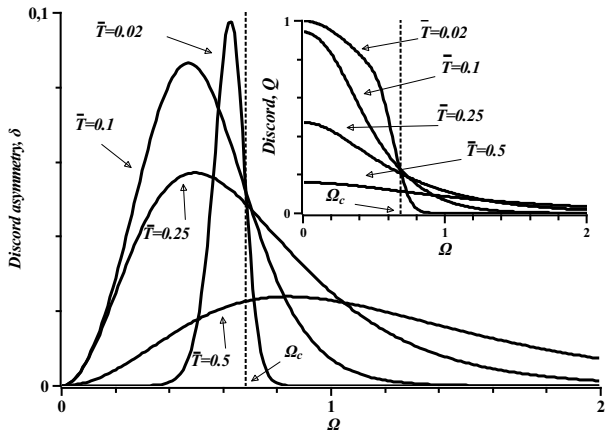
   , scale=0.5,angle=270
}
  \caption{Fig.3. The discord asymmetry   $\delta=Q^A-Q^B$ and the  discord $Q$  (inset) as functions of $\Omega$ for $\Delta=\Delta_{max}\approx0.683$ and different values of   the dimensionless  temperature $\bar T$: $\bar T=0.02,0.1,0.25,0.5$. The vertical dotted line corresponds to $\Omega=\Omega_c(\Delta_{max})=0.684$, see eq.(\ref{Omegac}).}
  \label{Fig:Om} 
\end{figure*}

We also analyze the discord asymmetry  $\delta$ as a function of $\Omega$ for the different values of the dimensionless temperature $\bar T=kT/(\hbar D)$ at $\Delta\approx 0.683$: $\bar T=0.02,0.1,0.25,0.5$, see Fig.\ref{Fig:Om}. It was mentioned above that the discord asymmetry of the ground state ($T=0$) appears only for $\Omega=\Omega_c$. On the contrary, if $T>0$ then  the asymmetry  exists for any $\Omega$ and  it becomes essential inside of some interval around the critical value $\Omega_c$. This interval increases with increase in the temperature, which is demonstrated in Fig.\ref{Fig:Om}. { Conversely, if $T\to 0$, then the asymmetry becomes localized in the neighborhood of  $\Omega_c$. The asymmetry reduces to the single point $\Omega=\Omega_c$ at $T=0$ (the ground state)}. 

The maximum of $\delta$ for $T=0.02$ is found at $\Omega\approx 0.497$, $\delta_{max}\approx 0.098$. It corresponds to $Q\approx 0.399$ so that the  asymmetry is about $25 \%$. Similarly, the maximal asymmetry at $\bar T=0.1,\;0.25, \;0.5$ is about $19 \%$,  $20 \%$ and $24 \%$  respectively. { We see that the relative asymmetry $\frac{\delta}{Q}$ remains significant inside of the whole temperature interval considered here although both $\delta$ and $Q$ are vanishing with the increase in $T$.}


\section{Conclusions}
\label{Section:conclusions}
We have studied the asymmetry of the quantum discord in the bipartite system for the both $T=0$ (the ground state) and  $T>0$ with different values of parameters in the Hamiltonian. 


The dependence of the discord on the choice of the subsystem for the projective measurements directly follows from  the dependence of the classical part of the mutual information  on this choice. 
Our study demonstrates  that, although the description of quantum correlations by the 
quantum discord is a  significant progress in comparison with such description by the entanglement,  it requires the further development.

This work is supported 
by the Program of the Presidium of RAS 
No.18 "Development of methods of obtaining chemical compounds and creation of new
materials".


\begin{thebibliography}{99}


\bibitem{W}
R.F.Werner, Phys.Rev.A {\bf 40}, 4277 (1989)

\bibitem{HW}
S.Hill and W.K.Wootters, Phys. Rev. Lett. {\bf 78}, 5022 (1997)



\bibitem{P}
A.Peres, Phys. Rev. Lett. {\bf 77}, 1413 (1996)


\bibitem{AFOV}
L.Amico, R.Fazio, A.Osterloh and V.Ventral, Rev. Mod. Phys. {\bf 80}, 517 
(2008)

\bibitem{DPF}
S.I.Doronin, A.N.Pyrkov and E.B.Fel'dman, JETP Letters {\bf 85}, 519 (2007)

\bibitem{BCJLPS}
S.L.Braunstein, C.M.Caves, R.Jozsa, N.Linden, S.Popescu and R.Schack, Phys.Rev.Lett. {\bf 83}, 1054 (1999)

\bibitem{M}
 D. A. Meyer, Phys. Rev. Lett. {\bf 85}, 2014 (2000).

\bibitem{LBAW}
B. P. Lanyon, M. Barbieri, M. P. Almeida, and A. G. White, Phys. Rev. Lett. {\bf 101}, 200501 (2008).



\bibitem{BDFMRSSW}
C.H.Bennett, D.P.DiVincenzo, C.A.Fuchs, T.Mor, E.Rains, P.W.Shor, J.A.Smolin and 
W.K.Wootters, Phys.Rev. A {\bf 59}, 1070 (1999)

\bibitem{OZ}
H.Ollivier and W.H.Zurek, Phys.Rev.Lett. {\bf 88}, 017901 (2002) 


\bibitem{L}
S.Luo, Phys.Rev.A {\bf 77},  042303 (2008)

\bibitem{ARA}
M.Ali, A.R.P.Rau and G.Alber, Phys.Rev.A {\bf 81}, 042105 (2010)

\bibitem{DSC}
A.Datta, A.Shaji and C.M.Caves, Phys.Rev.Lett. {\bf 100}, 050502 (2008)



\end{thebibliography}
\end{document}